# Time delay of slow electrons-endohedral elastic scattering

M. Ya. Amusia[a, b]* and A.S. Baltenkov[c]

[a]Racah Institute of Physics, the Hebrew University, Jerusalem, Israel;
[b]A. F. Ioffe Physical-Technical Institute, St. Petersburg, Russian Federation;
[c]Arifov Institute of Ion-Plasma and Laser Technologies, Tashkent, Uzbekistan

We discuss the temporal picture of electron collisions with fullerene. Within the framework of a Dirac bubble potential model for the fullerene shell, we calculate the time delay in slow-electron elastic scattering by it. It appeared that the time of transmission of an electron wave packet through the Dirac bubble potential sphere that simulates a real potential of the $C_{60}$ reaches up to $10^4$ attoseconds. Resonances in the time delays are due to the temporary trapping of electron into quasi-bound states before it leaves the interaction region. As concrete targets we choose almost ideally spherical endohedrals $C_{20}$, $C_{60}$, $C_{72}$, and $C_{80}$. We present dependences of time-delay upon collision energy.

**Introductory remarks**
The time-picture of processes with participation of microscopic particles was suggested relatively long ago, being a natural continuation of the classical description. However since than it existed in a dormant state, and was almost completele substituted by quantum-mechanical picture that can be called energy-picture. This picture for decades was considered as most suitable in description of the microwrld and dominated in articles and textbooks.

However, relatively recent experiments with atomic and molecular photoionization under the action of attosecond laser pulses have considerably altered the situation. Attosecond spectroscopy provides a remarkable possibility of insight into the process of the outgoing wave packet formation within the target. It permits to investigate the inner processes in the target that accompany its ionization. The investigation of photoionization by attoseconds lasers demonstrated that time-picture better suites to describe these processes. Indeed, they have provided the first observation of the time delay in photoemission (see e.g. review [1] and references therein). It has been found that the photoionization is not instantaneous: the departure of the photoelectron wave packet is temporarily delayed relative to the arrival of the electromagnetic pulse, typically by a few attoseconds. This physical effect was called as "time delay in *half-scattering process*", in the contrary to the time delay originally introduced by [2-4]. In these papers the time delay (that is referred to as the EWS-time delay) is due to capture and retain for some time a particle by short-range potential in the resonance *elastic scattering process*.

In our recent paper [5] within the framework of a Dirac bubble potential model we calculated the partial Wigner-times delay $\tau_l(E)$ in slow electron elastic scattering by the fullerene $C_{60}$. It has been shown that the time of transmission of an electron wave packet through the Dirac bubble potential sphere (that simulates a real potential of the $C_{60}$ cage) exceeds by more than an order of magnitude the transmission time via an atomic electron shell. Resonances in the time delays are due to the temporary trapping of electron into quasi-bound states before it leaves the potential sphere.

Here we extend the consideration of [5] to other very close to spherical symmetry endohedrals, namely $C_{20}$, $C_{72}$, and $C_{80}$.

**Essential formulas**

In the special case of potential scattering with spherical symmetry the EWS-time delay is simply the derivative of the particle scattering phase shift $\delta_l(E)$ with respect to particle kinetic energy $E$[1]

$$\tau_l(E) = 2\frac{d\delta_l(E)}{dE}. \tag{1}$$

The Dirac bubble potential model gives the following expression (see Eq. (10) in [6]) for $e$-$C_{60}$ elastic scattering phase shifts

$$\delta_l = \arctan\left[\frac{xj_l^2(x)}{xj_l(x)n_l(x) - 1/R\Delta L}\right]. \tag{2}$$

Here $j_l(x)$ and $n_l(x)$ are the spherical Bessel functions with the asymptotic behavior

$$j_l(x \to \infty) = \sin(x - \pi l/2)/x; \qquad n_l(x \to \infty) = -\cos(x - \pi l/2)/x, \tag{3}$$

where the variable is $x = kR$ with $k = \sqrt{2E}$ being the electron momentum in continuum. The parameter $\Delta L$ in formula (2) is a jump of the logarithmic derivative of the electron wave functions at $r = R$ where the potential $U(r) = -U_0\delta(r - R)$ is infinitely negative.

At the point $r = R$ the electron wave functions, for both bound and continuum states, are smooth but their derivatives experience a jump. The following expression $\Delta L = -2U_0$ connects this jump with the strength of the potential $U_0$. The experimental data on electron affinity ($EA$) of $C_{60}^-$ provides information on the $\Delta L$ value. If the extra electron in the $C_{60}^-$ ground state is an $s$-state, then $\Delta L$ is defined by the following expression (for details see [6])

$$\Delta L = \Delta L_{1s} = \left[\frac{d}{dr}\ln K_{1/2}(\beta r)\right]_{r=R} - \left[\frac{d}{dr}\ln I_{1/2}(\beta r)\right]_{r=R}. \tag{4}$$

For the extra electron in the $p$-ground state of $C_{60}^-$ instead of (4) we have

$$\Delta L = \Delta L_{2p} = \left[\frac{d}{dr}\ln K_{3/2}(\beta r)\right]_{r=R} - \left[\frac{d}{dr}\ln I_{3/2}(\beta r)\right]_{r=R}. \tag{5}$$

In formulas (4) and (5) the functions $I_\nu(\beta r)$ and $K_\nu(\beta r)$ are the modified spherical Bessel functions exponentially decreasing with the rise of $r$; $\beta$ is connected with $EA$ by relation $\beta = \sqrt{2I}$ where $I = -EA$.

Variation of the bubble potential parameters $U_0$ and $R$ opens up the opportunity to apply this model to simulate and explore in natural time scale the electron interaction not only with fullerene $C_{60}$ but also with a whole class of spherical carbon molecules. We use this opportunity in the present paper, performing calculations for fullerenes $C_{20}$, $C_{72}$ and $C_{80}$ in order to obtain for these objects the partial Wigner times delay.

---

[1] Here we use the atomic system of units $e = m = \hbar = 1$, where e, m are electron charge and mass, while $\hbar$ is the Planck constant.

Only two experimentally observed parameters, namely the fullerene radius $R$ and the affinity energy $EA$ of an extra electron to the neutral fullerene are the fitting parameters. The generalization of Dirac bubble potential model to the case of any near spherical fullerenes is elementary. To do this, we have to replace the parameters of this model in formulas for $C_{60}$ [6] according to the Table below, where all parameters are given in the atomic units.

**Table**. Fullerenes parameters in the Dirac bubble potential model

|  | Fullerene radius, $R$ | Affinity energy, $EA$ | Potential strength, $U_0$ | Jump of log-derivative, $\Delta L$ |
|---|---|---|---|---|
| $C_{60}$ [6] | 6.639 | -0.0974 | 0.443 | -0.885 |
| $C_{72}$ | 7.084 [7] | -0.1166 [8] | 0.483 | -0.967 |
| $C_{80}$ | 7.622 [7] | -0.1165 [8] | 0.483 | -0.966 |
| $C_{20}$ | 3.855 [9] | -0.0827 [10] | 0.425 | -0.850 |

The details of numerical calculations of phase shifts $\delta_l(E)$ and partial EWS-times delay $\tau_l(E)$ in the Dirac bubble potential model are given in paper [5] and we drop down them here.

**Results of calculations**

In the Figure we present the partial EWS-times delay $\tau_l(E)$ for $C_{20}$, $C_{60}$, $C_{72}$ and $C_{80}$ fullerenes as functions of the electron kinetic energy $E$. In the upper two panels EWS-times delay are negative at small electron energy. Consequently, the potential fields of all considered fullerenes are capable to support $s$- and $p$-discrete levels [11]. The results demonstrate strong sensitivity of the results obtained to the fullerenes size. Presence of a bound $l$ discrete level dictates the respective time-delay to be negative at $E \to 0$.

Close to zero $E$ the $s$-phase shift behaves as $\sim (\pi - E^{1/2})$ if a single negative ion bound $s$-state exists. Appearance of additional bound state leads to a jump in the time delay from + to – infinity at $E \to 0$. The time delay reaches its maximum values for $f$ and $g$-wave at $E=$ 0.05 and is there as big as 500 and 510 atomic units, respectively that is about $1.3 \times 10^4$ attoseconds. It appeared that $d$-discrete levels exist in potential wells of $C_{60}$, $C_{72}$ and $C_{80}$ fullerenes. Fullerene $C_{20}$ has no $d$-discrete level and therefore for it $\tau_2(E)$ for small values of $E$ is positive [11]. There are discrete $f$-levels in $C_{72}$ and $C_{80}$ fullerenes while $C_{60}$ and $C_{20}$ have not it. Therefore $\tau_3(E)$ for them is positive. However, for $C_{60}$ the potential well is close to supporting a discrete $f$-level that leads to strong positive shape resonance in the $f$-partial wave [6]. Similar resonances for $g$-waves EWS-times delay one can see in the lower left panel in the cases of $C_{60}$, $C_{72}$ and $C_{80}$. Resonances in $h$-EWS-times delay $\tau_5(E)$ for these fullerenes, but with considerably smaller amplitudes, present the lower right panel.

**Discussion and conclusions**

Of interest would be to experimentally observe the specific features of the EWS-times delay in scattering processes presented in the figure. However in the present days it is hardly possible. It does not mean that the calculations carried out here are a some kind of mathematical exercises. Such measurements are waiting for their time. In addition, note that the EWS-times delay in elastic electron scattering with an accuracy of a factor of two coincide with the times delay in *half-scattering processes*. So, EWS-time can be used in interpretation of ongoing experiments with photoionization processes.

Indeed, let us consider the photoionization of *np*-subshell of atom A located in the center of $C_{60}$ cage, thus forming an endohedral $A@C_{60}$. This location results in emission of photoelectrons that have to penetrate the $C_{60}$ shell. It is natural to assume that in the first approximation the time delay of photoelectron wave packet $\tau_l^{A@C60}$ is equal to the sum of two differ times delay

$$\tau_l^{A@C60} = \tau_l^A + \tau_l^{C60}. \qquad (6)$$

The first of these terms is associated with the passage of the wave packet through the field of the atomic residue. The second one is simply

$$\tau_l^{C60}(E) = \tau_l(E)/2, \qquad (7)$$

that is one half of EWS-times delay presented in figure 1. So, the difference between departure of *s*- and *d*-electrons depend on, including, the calculated here EWS-times $\tau_0(E)$ and $\tau_2(E)$. Such time delay for pure atomic targets was observed and measured in recent experiments [12].

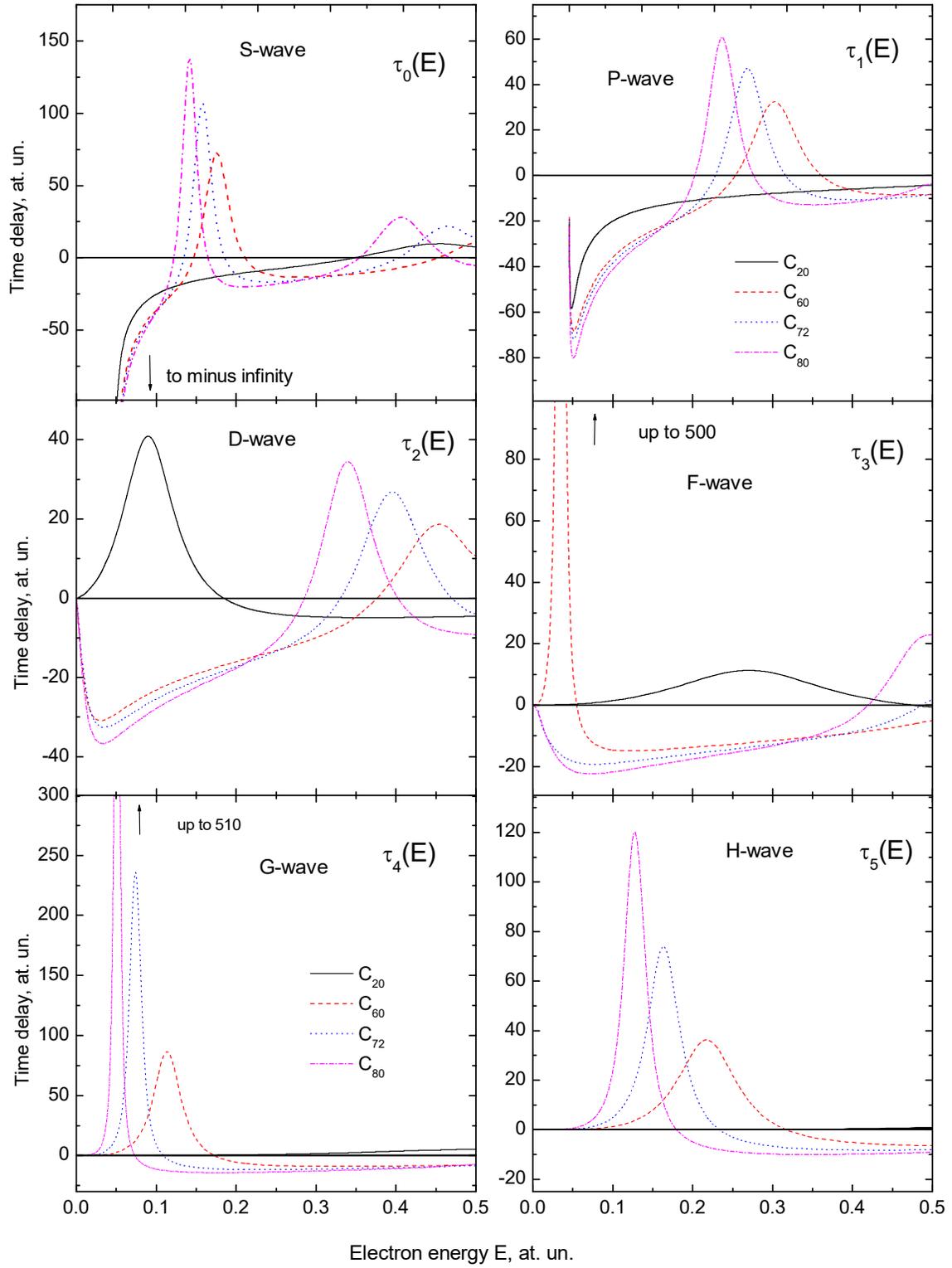

Figure 1. Partial EWS-times delay $\tau_l(E)$ of $C_{20}$ (solid black line), $C_{60}$ (dashed red line), $C_{72}$ (dotted blue line) and $C_{80}$ (dash-dotted magenta line) fullerenes as functions of electron kinetic energy $E$.

5